\documentclass[aps]{revtex4} 

\usepackage{epsfig}
\usepackage{syntonly} 

\newcommand{\omm}{\Omega_{m0}}

\newcommand{\bc}{\begin{center}}
\newcommand{\ec}{\end{center}}
\newcommand{\rat}{\mathcal R_0}
\newcommand{\dub}{w_Q}

\newcommand{\rhoq}{\rho_Q}

\newcommand{\be}{\begin{equation}} 
\newcommand{\ee}{\end{equation}}
\newcommand{\bear}{\begin{eqnarray}}
\newcommand{\eear}{\end{eqnarray}}
\newcommand{\bitem}{\begin{itemize} \setlength{\itemsep 3pt} }
  \newcommand{\eitem}{\end{itemize}}
\newcommand{\benum}{\begin{enumerate} \setlength{\itemsep 3pt} }
  \newcommand{\eenum}{\end{enumerate}}

\begin{document}

\title{Observed Smooth Energy Fitted by Parametrized Quintessence} 
\author{S. A. Bludman}\email[]{bludman@mail.desy.de}
\affiliation{Deutsches Elektronen-Synchrotron DESY,
  Hamburg\\ University of Pennsylvania, Philadelphia}
\date{\today}
\begin{abstract}
This paper proposes two phenomenological quintessence potentials with
parameters fitted to the presently observed ratio $\rat$ of smooth
energy to clustered mass and limits on the equation of state parameter
$w_Q(a)=P/\rhoq$, for which the quintessence potential is now slow- or
fast-rolling. These two potentials are intended
to illustrate rather different quintessence phenomenology since
tracking and into the near future. Neither can be a
fundamental potential derivable from present-day string theory.
\end{abstract}

\maketitle

The condition for a tracker solution to exist is \cite{SWZ} that the
logarithmic derivative of the scalar field potential, $-V'/V,~V'\equiv
dV/d\phi$, be a slowly decreasing function of scalar field $\phi$ or
that 
\be 
0<\Gamma(a)-1 \equiv d(-V/V')/d\phi 
\ee 
be nearly constant as
function of cosmological scale $a$. It is convenient to measure the
steepness of the potential by the logarithmic derivative $\beta(\phi)
\equiv -d\ln V/d\ln\phi$, so that \be
\Gamma(a)-1=d(\phi/\beta)/d\phi=1/\beta(\phi)\cdot(1-d\ln\beta/d\ln\phi).
\ee We will consider below, an inverse-power potential, for which
$\beta$ is strictly constant, and an ``isothermal'' potential, for
which $\beta$ slowly decreases with $\phi$ so as to keep the equation
of state $\dub=constant$ \cite{cosmo8}.
 
The ordinary inflationary parameters are then \be\eta(\phi)
\equiv V''/V ,~~~~2\epsilon\equiv (V'/V)^2, ~~~~\Gamma=V\ddot
V/\dot V^2=\eta(\phi)/2\epsilon(\phi).\ee  When
$\eta,~2\epsilon\ll 1$, a tracking potential will be {\em
slow-rolling}, meaning that $\ddot\phi$ in the scalar field equation of motion
and $\dot\phi^2$ in the quintessence energy density are both negligible.  This slow-roll
approximation is usually satisfied in ordinary inflation.  In the
early e-folds of tracking quintessence, however,
$-V'/V=\beta(\phi)/\phi$ is slowly changing, but need not itself be
small, if $\beta \neq 0$.  This establishes the important
distinction between static or quasi-static quintessence, for which the
EOS parameter and scale exponent $n_Q(a)=3(1+w_Q)$
have present values $n_Q(1)\sim 0,~w_{Q0}\sim -1$, and sensibly
dynamical quintessence, for which
$n_Q>1,~w_{Q0}>-2/3$.  At the observationally allowed upper limit
$w_{Q0}=-0.5$, the first few scale e-folds would still be {\em
  fast-rolling}. Such dynamical quintessence will become slow-roll
only in the very far future, after many e-folds of the
quintessence field slowly driving $\Delta$ from $1$ to zero and
$\dub$ from its present value to zero.  This means that while the
slow-roll approximation is applicable to ordinary inflation,
dynamical quintessence generally requires exact solution of the equations
of motion.

For tracking potentials derived from physical principles, it is
reasonable to assume that $\beta \sim constant$, at least since
tracking began.  We therefore parametrize quintessence so that the
SUSY-inspired potential \be V(\phi)=V_0(\phi_0/\phi)^\beta,~~~~
\beta=3.5 \ee is inverse power law, and $\Gamma
-1=1/\beta=constant$.

For comparison, we also consider the isothermal equation of state
\be
V(\phi)=V_0[\sinh(\alpha\phi_0)/\sinh(\alpha\phi)]^{\beta},\quad
\beta=2, \quad \alpha \equiv \sqrt{3/(2\beta+\beta^2)}=0.612 ,
\ee 
for which $\Gamma -1=\Omega_B/\beta$ is not constant, but
decreases from $1/\beta$ when tracking begins, to zero in the far
future, as the clustered mass fraction $\Omega_B->0 .$ This potential
is called ``isothermal'' because for it, the quintessence
pressure/density ratio 
$\dub= -2/(2+\beta)=-1/2$ and $n_Q=3\beta/(2+\beta)=3/2$ are
constant once matter dominates over radiation. This makes the tracker
approximation exact for the
isothermal potential: $\Delta=1$
and $-V'/V=n_Q \cdot da/a$. This
potential allows a scaling solution $V(a),~\rho_Q(a) \sim a^{-n_Q}$
\cite{Matos} and has the inverse power potential
$V=V_0(\phi_0/\phi)^{\beta}$ and the exponential potential
$V=V_0\exp(-\sqrt{n_{Q}} \phi)$ as limits for $\alpha\phi\ll 1$
and $\gg 1$, respectively.  It therefore interpolates between an
inverse power potential at early times and an exponential
potential at late times.

In each of these potentials, the two parameters have been chosen to
fit the present values $\rat=2$ and
$\rho_{Q0}=2\rho_{cr}/3=2.7E-47~GeV^4$, so that for $w_{Q0}=-0.5$,
$n_{Q0}=3/2$ the present value of the potential is
$V_0=V(\phi_0)=1.013E-47~GeV^4$. This requires tracking to begin 
after matter dominance,  reaching present values $\phi_0=2.47$ for
the inverse-power and $1.87$ for the isothermal EOS, respectively.
Tracking continues indefinitely for the isothermal EOS, but is of
relatively short duration for the inverse-power potential.
We believe these parametrizations of $V(\phi)$ to be representative of
reasonable smooth potentials, over the red-shift range $z<5$
that is observationally accessible. If we considered lower values of
$\beta$ for the present ratio $\rat=2$, $w_{Q0}$ would then decrease from
$-1/2$ to $-1$ as the potential became faster rolling.

For the inverse-power potential, $-V'/V= \beta/\phi$, and for the
isothermal potential $-V'/V=\beta\alpha/\tanh(\alpha\phi)$
respectively, so that, for $w_{Q0}=-0.5$, $\eta$ and perforce
$\epsilon$ are not now small.  Thus both the dynamical quintessence
potentials we are considering are now still fast-rolling.  The
isothermal potential, with $\beta=2$, asymptotically approaches
$~\exp(-1.22~\phi)$ and will never be slow-rolling.  The inverse-power
potential, with $\beta=3.5$, will become slow-rolling once
$\phi>\beta$ and will asymptotically approach a de Sitter solution in
the distant future. In the observable recent past, its $\dub$
increases with $z$ approximately as $\dub(z)\approx-0.5+0.016z$.

After transients depending on initial conditions, and a frozen
epoch ($\dub\approx -1$), the scalar field overshoots and then
converges onto a solution \be
n_Q(a)=n_B(a)\beta/(2+\beta)=n_B(a)/2\ee
which tracks the
background $n_B(a)$. During the tracking regime, until
quintessence dominates, $\Delta\approx 1$ and $\Omega_Q\approx
n_Q\phi^2/\beta(\phi)$ increases quadratically with field
strength.
\begin{figure*}  
\epsfig{figure=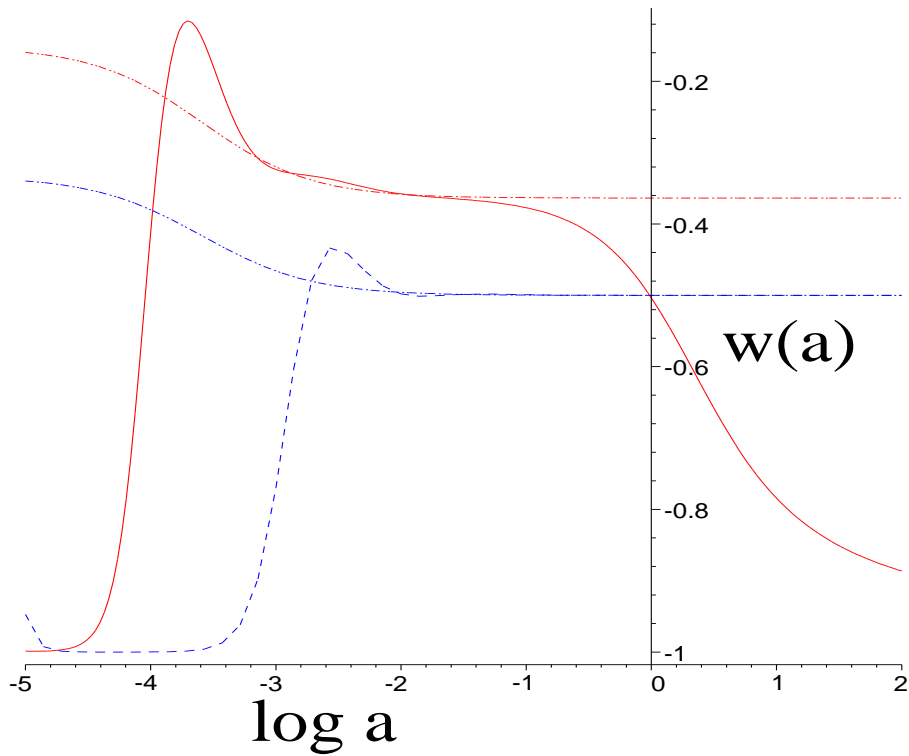,height=10cm, width=12cm}
\caption{Exact and tracker approximation solutions $\dub$ for inverse power and
  isothermal potentials chosen to give present
  values $\rat=2,~~w_{Q0}=-0.5$.  The exact $\dub$ values for the
  inverse-power and the isothermal quintessence potentials (4) and (5)
  are shown by the solid and dashed curves. The tracker approximation
  (dash-dotted curves) is exact for the isothermal potential, but holds
  only briefly for the inverse-power potential and considerably
  overestimates $\dub$ once quintessence is appreciable.
  Asymptotically, $\dub->-0.5,~1$, so that, if unchanged,
  both potentials would show an event horizon.}
\end{figure*}
Driven by the background EOS which is changing around $z_{cr}=3880$,
the quintessence equation of state parameter $\dub(a)$
slowly decreases. The inverse-power potential reaches its tracker at
$\log a \approx -3.2$, but remains there only until $\log a \approx
-1.8$, when the growth of the quintessence field slowly drives $\dub$
down from the tracker value $-0.364$ towards $-1$ in the very far
future (solid curve in Fig. 1).  For the inverse-power potential, the
tracker approximation holds only briefly, and thereafter seriously
overestimates $\dub$. The exact isothermal solution reaches
its tracker later, at $\log a >-1$, but remains exactly on tracker
thereafter (dashed curve in Fig. 1).

Once reaching the tracker, the isothermal  equation of state
stays constant at $\dub(\phi)=-2/(2+\beta)=-1/2$.
The evolution with cosmological scale is fixed by the
scaling $V/V_0=\rho_Q/\rho_{Q0}=a^{-3/2}$, which makes 
\be
\mathcal R(a)=\sinh^2 (\alpha \phi),\quad \Omega_Q=1-\Omega_B=\tanh^2
(\alpha \phi)=\rat a^{3/2}/(1+\rat a^{3/2}) .
\ee 
For the present ratio $\rat \sim 2$ one has $\alpha \phi_0\leq
1.146$ and $\phi_0 \leq 1.87$. The observations that the Universe
is now accelerating fix the bounds  
\be
-2/(2+\beta) \leq -1/2,~~~\beta \leq 2, ~~~\alpha\geq
0.612,~~~\phi_0 \leq 1.87. 
\ee 
For $\beta=0$, the potential is
static: $\Lambda$-dominance started at redshift
$R_0^{1/3}-1=0.260$, the universe was then already accelerating
$-q=0.333$, and is now accelerating faster $-q_0=0.5$.  For
larger $\beta$, the scalar field is more dynamic.  At the upper
limit, $\beta=2$ the isothermal potential rolls only as fast as
the observed acceleration ($-q_0 \geq 0$) allows, tracking starts
only after matter domination, and acceleration starts only now.
Nevertheless, $\rho_Q(a)$ was still subdominant to the radiation
density all the way back into the era of Big Bang nucleosynthesis
\cite{Bean}. 

As mentioned above, the tracking approximation is exact for the
isothermal potential, but only briefly valid for the inverse-power potential.
For $\alpha\phi\ll 1$, the background dominates and the isothermal
equation of state reduces to the inverse power potential $V\sim
\phi^{-\beta}$, for which, along the tracker, $\phi,~\sqrt{\mathcal
  R}\sim a^{3/(2+\beta)}$. We shall, however, also be interested in
the quintessence dominated era $\alpha\phi>1$, when different
potentials give different present and future behavior.

If $\dub$ is not constant, we could still reconstruct the quintessence
potential from $\dub(z)$.  Indeed $d \dub(z)/dz$ is positive
generically for quintessence and generally negative for k-essence
\cite{Picon}, an alternative in which the scalar field has a
non-linear kinetic energy instead of a potential.  In principle,
$\dub(z)$ is observable in high red-shift supernovae \cite{SNAP}, in
cluster evolution \cite{Newman,Haiman} and in gravitational lensing
\cite{Cooray}.  In practice, $\dub(z)$ is poorly constrained
observationally and theoretically \cite{Barger,Limitations,Astier}.
Over the small redshift range for which smooth energy dominates and
for which sensitive measurements are possible, the effects of varying
$\dub(z)$ are much smaller than the present uncertainties in
measurement and in cosmological model.  Theoretically, the luminosity
distance to be measured in high-redshift supernova and in comoving
volume number density measurements depends on integrals which smooth
out the sensitivity to $\dub(z)$ \cite{Limitations}.  Until $\omm$ is
determined to (1-2)\% accuracy, SNAP \cite{SNAP} and other future
experiments will only be able to determine some $\dub$ effective over
the observable redshift range $z<2$, to tell whether the smooth energy
is static or dynamic.  In the far future, however, the differences among
different quintessence potentials will asymptotically become
substantial.

The difference between the two potentials (4,5) appears only in the
evolution of $\dub$ now and in the future, when $\alpha\phi>1$.  For
the inverse power potential, the growth of quintessence drives
$\Delta(a)$ below unity and $w_Q(a)$ decreases slowly towards -1. The
quintessence energy density $\rhoq$ and the expansion rate
$H\rightarrow constant$, drive ultimately towards a de Sitter
universe, which inflates exponentially.  The isothermal EOS will
approach the exponential potential, $\dub$ stays constant, $\rhoq$ and
$H$ continue to decrease, and inflation is power-law.  For both these
potentials, acceleration continues indefinitely, so that two observers
separated by fixed coordinate distance, ultimately have relative speed
$\geq c$.  Because an event horizon exists in both cases, a local observer
cannot construct an S-matrix.  This shows that neither of our
phenomenological potentials is derivable from field theory or string
theory. To be string-inspired and lead to an event horizon, the
quintessence EOS would have to ultimately change back from
accelerating to decellerating ($\dub>-1/3$) \cite{He}.

This paper derives from joint work with Matts Roos \cite{cosmo8}.


\end{document}